\documentclass[pdflatex,sn-mathphys,iicol]{sn-jnl}
\usepackage{isotope}

\jyear{2021}%

\theoremstyle{thmstyleone}%
\theoremstyle{thmstyletwo}%

\theoremstyle{thmstylethree}%

\begin{document}

\title[Article Title]{Towards a Unified Model of Neutrino-Nucleus Interactions}

\author[1]{\fnm{Omar} \sur{Benhar}}\email{omar.benhar@roma1.infn.it}

\author[2]{\fnm{Camillo} \sur{Mariani}}\email{camillo@vt.edu}

\affil[1]{\orgdiv{INFN}, \orgname{Sezione di Roma}, \orgaddress{ \city{00185 Roma}, \country{Italy}}}
\affil[2]{\orgdiv{Center for Neutrino Physics}, \orgname{Virginia Tech}, \orgaddress{ \city{Blacksburg}, \state{Virginia 24060}, \country{USA}}}


\abstract{The achievement of the goals of the ongoing and future accelerator-based neutrino 
experiments\textemdash notably the determination of CP violating phase in the lepton sector\textemdash will require
the development of advanced models of neutrino-nucleus interactions. In this review, we summarise the present status
of experimental studies of neutrino-nucleus scattering, and discuss the developments and perspectives of the theoretical approach
based on factorisation of the nuclear cross section, which has recently emerged as a promising framework for the description of the variety of processes contributing to the detected signals.}


\maketitle
\section{Introduction}  
\label{intro}

Over the past two decades, theoretical and experimental investigations 
of lepton interactions with complex nuclei\textemdash such as carbon, oxygen and 
argon\textemdash have emerged as a prominent area of intersection between 
particle and nuclear physics.

A fully quantitative understanding of neutrino-nucleus interactions at energies up to few GeV 
is needed for the interpretation of the data collected by accelerator-based studies of neutrino 
oscillations. The achievement of this goal, however, involves non-trivial problems, and requires the development of advanced cross section models, suitable to reliably describe the complexity of nuclear dynamics. Such models should also be computationally efficient, to satisfy the need of neutrino interaction generators.

The main difficulty involved in the analysis of the measured neutrino-nucleus cross sections
stems from the fact that the beam energy is distributed over a broad spectrum. As a consequence, 
the observed signals receive contributions from a variety of competing reaction mechanisms\textemdash 
ranging from quasi elastic scattering to resonance production and deep inelastic scattering\textemdash whose clear 
cut identification requires a consistent description of all the underlying microscopic processes.  

The growing realisation that the relativistic Fermi gas model (RFGM)\textemdash routinely employed in Monte Carlo simulations of neutrino interactions\textemdash
fails to explain the available data has triggered a great deal of effort to develop more realistic 
models, some of which have arguably reached the degree of maturity 
required for a meaningful comparison between their predictions and the measured cross 
sections~\cite{BFNSS,Leitner:2008ue,Benhar:2010nx,Martini2010,nieves,Ankowski2013,NoemiPRL,SuSa,natalie}.

In spite of the different nature of the beam-target interaction, the large body of precise electron scattering data, 
and the theoretical methods that have been successfully employed for their analysis, provide 
much-needed guidance towards the development of an accurate treatment of the neutrino-nucleus
cross section. The ability to describe  electromagnetic processes is, in fact, an obvious  prerequisite to be met by 
any models of the nuclear response to weak interactions.

The formalism based on factorisation of the nuclear cross section, allowing to clearly separate the lepton-nucleon 
interaction vertex from the amplitudes involving the nuclear initial and final states, appears to be ideally suited to highlight and exploit the similarities between processes involving electrons and neutrinos. A detailed discussion of the emergence of factorisation can be found in, e.g., Refs.~\cite{BFNSS,Benhar:2006wy}. 

Factorisation clearly emerges in the kinematic region of high momentum transfer, in which the beam 
particles predominantly interact with individual nucleons\textemdash with the remaining target constituents acting as spectators\textemdash and the impulse approximation (IA) is applicable. 
Ample experimental evidence of the onset of the IA regime is provided by the observation 
of $y$-scaling in inclusive electron-nucleus scattering~\cite{sick80,Donnelly_superscaling}. 

Within the IA, lepton-nucleus scattering reduces to the incoherent sum of 
elementary processes involving off-shell nucleons, whose momentum and energy distribution is described by the target spectral function~\cite{PKE,LDA}. The key advantage of the spectral function formalism 
is the capability to combine an accurate description of nuclear dynamics, that can be obtained from non relativistic many-body theory, with a fully 
relativistic treatment of the interaction vertex, which is an absolute requirement 
when it comes to describing interactions of accelerator or atmospheric neutrinos.
This scheme, has been remarkably successful in explaining electron-nucleus scattering data spanning a broad kinematic range, and its extension to the treatment of weak interactions does not involve any conceptual difficulties.
Over the past decade, the factorisation {\it ansatz}\textemdash extended to allow a consistent treatment of 
processes in which the beam particle interacts with two nucleons\textemdash has established 
itself as a viable and promising scheme to analyse the flux-averaged neutrino-nucleus cross section.

It should be pointed out that in the independent particle model of the nucleus\textemdash the simplest 
implementation of which is the RFGM~\cite{RFGM}\textemdash factorisation is realised by construction, 
regardless of momentum transfer. However, a comparison to electron scattering data unambiguously shows that
the spectral functions derived within this scheme conspicuously fail to capture important features of nuclear dynamics. More advanced models, explicitly taking into account the effects of correlations among 
nucleons, have been developed by combining results of theoretical calculations and experimental information 
obtained from the measured $A(e,e^\prime p)$ cross sections. This approach has been successfully applied to the analysis of the electron scattering cross sections of medium-heavy nuclei, ranging between carbon and iron~\cite{LDA}, 
and has been recently extended to argon and titanium~\cite{PhysRevD.105.112002,https://doi.org/10.48550/arxiv.2209.14108}. The availability of the argon and titanium spectral 
functions will be critical to the description of neutrino and antineutrino interactions in the liquid argon detectors that will be employed by next-generation neutrino experiments, notably the Deep Underground Neutrino Experiment (DUNE) in the USA. 

The primary goal of this review is providing a summary of the current status 
of experimental studies of neutrino-nucleus interactions, and 
a critical discussion of the theoretical framework based on 
factorisation and the spectral function formalism. In this context, it is important to realise that, 
besides the approach specifically referred to as spectral function method, most models employed 
to study neutrino-nucleus cross sections\textemdash from RFGM~\cite{RFGM} to superscaling~\cite{Donnelly_superscaling}
and the recently proposed short-time approximation of Quantum Monte Carlo~\cite{PhysRevC.101.044612}\textemdash exploit some degree of factorisation. A notable exception is the approach based on the
Green's Function Monte Carlo technique~\cite{PhysRevC.91.062501}, which allows 
to carry out very accurate calculations of the electromagnetic and
weak responses of nuclei as heavy as carbon. Owing to its inherently non-factorisable 
nature, however, the applicability of this approach is limited to the kinematic region in which the non relativistic approximation is expected to be valid.

The review is organised as follows. In Section~\ref{oscillations} we outline the main elements of the analysis leading to the experimental determination of neutrino oscillations, while  Section~\ref{nuA:expt} provides a summary of the available measurements of the neutrino-nucleus cross sections.
The theoretical treatment of lepton-nucleus scattering in the IA regime, based on factorisation and the spectral function formalism, is discussed in Section~\ref{nuA:th}. 
The issues associated with the identification of 
the different reaction mechanism are analysed in Section~\ref{flux:average}. Finally, in Section~\ref{summary} we state the main conclusions emerging from our discussion, and outline the prospects for future work.

\section{Experimental determination of neutrino oscillations}
\label{oscillations}


Neutrino oscillation physics has become a precise science thanks to the advent of new accelerators and powerful beams. New oscillation experiments require the systematic errors to be understood at better than few percent level. Neutrino oscillation analyses are, in principle, quite simple: the experiment has to count the number of neutrino un-oscillated events at the source and then the number of oscillated events at the far detector. The ratio of these two quantities is directly related to the probability of neutrino oscillation, which, in the simplified two-neutrino framework, can be written as
\begin{eqnarray}
\frac{R^{Far}_{\alpha\rightarrow\beta}(\mathcal X)}{R^{Near}_{\alpha}(\mathcal X)} = P_{\alpha \rightarrow \beta} = \sin^2(2\theta)sin^2\left(\frac{\Delta m^2 L}{4E}\right).
\label{eq:nu_energy_dep}
\end{eqnarray}

In this formalism, the number of events at the near and far detectors, $R(\mathcal X)$, is described with respect to a set of observables that depend on the neutrino energy and, only in the case of the far detector, from the oscillation probability
\begin{align}
\label{eq:rate}
R^{Near}_{\alpha} (\mathcal X) = \mathcal N\int dE_\nu\Phi_\alpha(E_\nu)\frac{d\sigma_\alpha}{d\mathcal X}\epsilon_\alpha(\mathcal X) \\
R^{Far}_{\alpha\rightarrow\beta} (\mathcal X) = 
\mathcal N\int dE_\nu\Phi_\alpha(E_\nu)P(\nu_\alpha\rightarrow\nu_\beta)\frac{d\sigma_\beta}{d\mathcal X}\epsilon_\beta(\mathcal X)
\end{align}
In the above equations, $\mathcal X$ indicates the kinematic while the normalization $\mathcal N$ depends on the beam power, detector fiducial mass etc. The neutrino flux, $\Phi_\alpha(E_\nu)$, and the neutrino oscillation probability $P(\nu_\alpha\rightarrow\nu_\beta)$ are functions of the true neutrino energy $E_\nu$. The differential cross section ${d\sigma}/{d\mathcal X}$ describes the likelihood for a neutrino of given flavor ($\mathcal \alpha$ or $\mathcal \beta$) and energy $E_\nu$ to produce an event of kinematics $\mathcal X$, and  depends on the reconstructed neutrino energy. The last terms, $\mathcal \epsilon_\alpha$ and $\mathcal \epsilon_\beta$, represent the detection efficiency, and are functions of reconstructed neutrino energy and neutrino interaction types. Depending on the interaction that occurs at the vertex, different particles will be produced in final state, and the detector efficiency can be very different depending on the type of particle being detected; e.g. charged or neutral.

Conventional accelerator neutrino and antineutrino beams are produced using pion and kaon decay in flight. This technique relies on a 
magnetic horn to select the right pion and kaons and reduce the wrong sign particles. Flipping the polarity of the horn will change the beam from neutrino to anti-neutrino and vice-versa. The resulting flux is not monochromatic, its distribution in energy being usually broad. 
The mean energies of the neutrino and antineutrino beams are also different, and depend on the polarity of the horn. As a consequence, the neutrino and antineutrino beams are essentially different and correspond to independent experiments. Flux uncertainties originating from meson 
production data are a difficult issue to keep under control. In the most accurate studies\textemdash performed by the 
MINOS~\cite{MINOS:2009ugl} and MINER$\nu$A~\cite{MINERvA:2014ani} experiments\textemdash the level achieved is about 5\%, and it is based on a multitude of measurements, performed using multiple magnetic horn configurations and distances or varying the distances between horn systems and the target.

Understanding the cross sections and the associated nuclear models is also crucial for oscillation analyses, and has a direct impact on systematics. The required accuracy level should be based on the ability to measure the CP violating phase, which is obtained from the CP asymmetry
\begin{equation}
\label{eq:cpa}
A=\frac{\langle P\rangle-\langle \bar P\rangle}{\langle P\rangle+\langle \bar P\rangle}\,,
\end{equation}
where $\langle P\rangle$ is the energy averaged $\nu_\mu\rightarrow\nu_e$ oscillation probability and $\bar P$ is the corresponding quantity for antineutrinos. 

Performing accurate measurements of neutrino oscillation parameters in the presence of significant cross section and/or flux uncertainties is a long standing problem, and various experiments have used the ratio near to far detector to effectively cancel out systematics. In the reactor neutrino experiments, like the Daya Bay or DoubleChooz
experiments, for example, this ratio was used very effectively to measure the mixing angle $\theta_{13}$~\cite{An:2012eh}. For 
the cancellation to be effective, however, it is critical that the near and the far detectors have as close as possible responses to neutrino interactions, and in case there are any differences, they  have to be understood to high accuracy. Differences originate from many sources: detector response functions, geometric acceptance and/or different background levels. Recently, the T2K~\cite{T2K:2021xwb} and Nova~\cite{NOvA:2021nfi} Collaborations reported oscillation results on the $\nu_e$ cross section in a predominantly $\nu_\mu$ beam. The total systematic error is 
about 5-7\%, mostly coming from the beam flux uncertainty and the detector response. There are no obvious methods to improve this situation, 
and it will be worse in the case of antineutrino beams. In general, the contamination of wrong sign neutrino in a neutrino beam is of the order of few percent, but for wrong sign neutrino in an antineutrino beam this figure can be as 
large as 10\%. In general, the ability to measure neutrino and antineutrino cross section with much better accuracy than 
the one of the beam flux is known appears a difficult task, although new techniques like the T2K $\nu$PRISM concept~\cite{Scott:2015tsa} which utilizes data taken at several near detector off-axis locations, look quite promising.


\section{Measurement of the neutrino-nucleus cross section}
\label{nuA:expt}

Neutrino cross sections are a fundamental ingredient in neutrino oscillation analysis, and, to these days, they account for the bulk of the systematic uncertainties of neutrino experiments. In the last two decades neutrino scattering data have been collected with a multitude of beams and targets. Different analysis techniques have been developed and used in neutrino experiments providing an healthy data set, but it has not been always easy to compare measured neutrino cross sections on the same target from different experiments. This difficulty depends not only on flux or intrinsic beam energy spreads, or on the experimental techniques used, but also on the fact that neutrino scattering experiments do not always describe the cross section in term of fundamental observables, like momentum or energy. Sometimes, derived quantities like the squared four-momentum transfer Q$^2$ are also used, which makes the interpretation of the measurements difficult. Different experiments use varying assumptions\textemdash especially on how to subtract and evaluate backgrounds\textemdash and multiple neutrino interaction generator configurations to estimate backgrounds, detector efficiencies and nuclear effects. Here, we provide a summary of the results obtained in the last two decades by various experiments, with an emphasis on inclusive cross sections, the pion-less channel, also known as quasi elastic, and the one-pion production channel. For a more comprehensive review we refer the reader to Ref.~\cite{Formaggio:2012cpf}. 

\begin{table*}
\centering
\label{tab:xsection_results} 
\caption{List of neutrino accelerator experiments ($Exp$) and their measurements for neutrino and anti-neutrino inclusive and differential cross section ($Meas$), axial mass measured values ($M_A$), cross section with charged pion in final state ($\pi^{\pm}$) for both CC and NC reactions, cross section with neutral pion in final state ($\pi^{\circ}$) for both CC and NC reactions, nuclear targets and beam properties for both the neutrino and antineutrino data.}
\begin{tabular}{c c c c c c c }
\hline \hline
Exp.  & Meas. & $M_A$ & $\pi^\pm$ & $\pi^{\circ}$ & Target & $E^{Avg}_{\nu}$-$E^{Avg}_{\bar{\nu}}$  \\
\hline
&&&&& \\
NOMAD       & $\nu_{inc}$\cite{NOMAD:2007krq} & 1.0\cite{NOMAD:2009qmu} & - & NC~\cite{NOMAD:2009idt}     & C     & 23.4-19.7   \\
&&&&&\\
K2K       & - & 1.2\cite{K2K:2006odf} & CC\cite{K2K:2008tus,K2K:2005uiu} & CC\cite{K2K:2010xeb}     & CH      & 1.3- \\
          & - &  &  & NC\cite{K2K:2004qpv}     &   H$_2$O    &    \\
&&&&& \\
MiniBooNE & $\frac{d^2\sigma}{dT_{\mu}d\theta_{\mu}}$\cite{MiniBooNE:2010bsu,MiniBooNE:2013qnd}    & 1.2\cite{MiniBooNE:2007iti} & CC\cite{MiniBooNE:2009koj,MiniBooNE:2010eis} & CC\cite{MiniBooNE:2010cxl}     & CH$_2$      & 0.8-0.7 \\
          & NC\cite{MiniBooNE:2010xqw,MiniBooNE:2013dds}    &  & & NC\cite{MiniBooNE:2009dxl,MiniBooNE:2008mmr}     &      & \\
&&&&&& \\
MINOS & $\nu_{inc}$\cite{MINOS:2009ugl}, $\bar{\nu}_{inc}$\cite{MINOS:2009ugl}    & 1.2\cite{MINOS:2014axb} & - & -     & Fe      & 3.5-6.1 \\
          & &  & & NC\cite{MINOS:2016yyz}     &       \\
&&&&& \\
SciBooNE & $\nu_{inc}$\cite{SciBooNE:2010slc}    &  & CC~\cite{SciBooNE:2008bzb} & -     & CH      & 0.8-0.7 \\
          & &  & & NC\cite{SciBooNE:2009nlf,SciBooNE:2010lca}     &       \\
&&&&& \\
MINER$\nu$A & $\nu_{inc}$\cite{MINERvA:2015ydy,MINERvA:2020zzv,MINERvA:2021owq,MINERvA:2014rdw,MINERvA:2016oql,MINERvA:2016ing}, & & CC\cite{MINERvA:2014ogb,MINERvA:2016sfc,MINERvA:2014ani,MINERvA:2019rhx,MINERvA:2017ipy} & CC\cite{MINERvA:2015slz,MINERvA:2017okh,MINERvA:2020anu}     & He,C      & 3.5(LE) \\
& $\bar{\nu}_{inc}$\cite{MINERvA:2016ing}    &  & & NC\cite{MINERvA:2016uck}     & CH,H$_2$   & 5.5(ME) \\
            & $\frac{\nu_{inc}}{\bar{\nu}_{inc}}$\cite{MINERvA:2021wjs,MINERvA:2017ozn} &  & & & Fe&  \\
            & $\frac{d^2\sigma}{dp_Tdp_{\parallel}}$\cite{MINERvA:2018hqn,MINERvA:2018vjb,MINERvA:2019gsf}, &&&&Pb& \\
            & $\frac{d\sigma}{d\delta p_{Tx}}\frac{d\sigma}{d\delta p_{Ty}}$\cite{MINERvA:2019ope}, &&&&&  \\
            & $\frac{d\sigma}{dp_n}\frac{d\sigma}{d\delta\alpha_T}$\cite{MINERvA:2018hba}, &&&&& \\ 
            & $\frac{d\sigma}{dQ^2}$\cite{MINERvA:2013kdn,MINERvA:2013bcy,MINERvA:2017dzh}, &&&&& \\
            & 1p\cite{MINERvA:2014ypj}, &  & &      &    &   \\
            & $\frac{d^2\sigma}{dEdq_3}$\cite{MINERvA:2018nab},$\nu_e$\cite{MINERvA:2015jih} &  & &      &    &   \\
            & QE\cite{MINERvA:2022mnw} &  & &      &    &   \\
&&&&& \\
ArgoNeut & $\nu_{inc}$\cite{ArgoNeuT:2011bms,ArgoNeuT:2014rlj} & & CC\cite{ArgoNeuT:2014uwh,ArgoNeuT:2018und} & -     & Ar      & 4.3-3.6 \\
& $\bar{\nu}_{inc}$\cite{ArgoNeuT:2014rlj}    &&& NC\cite{ArgoNeuT:2015ldo} &       \\
& 2p\cite{ArgoNeuT:2014ihi}    &&&& &       \\
&&&&& \\
MicroBooNE & $\nu_{\mu}$\cite{MicroBooNE:2021sfa,MicroBooNE:2019nio,MicroBooNE:2018xad} & & CC\cite{MicroBooNE:2018neo} & CC\cite{MicroBooNE:2021ccs}     & Ar      & 0.8-0.7 \\
& $\nu_{e}$\cite{MicroBooNE:2022hun,MicroBooNE:2021gfj}   &&& NC\cite{MicroBooNE:2022lvx} &       \\
& $\bar{\nu}_{e}$\cite{MicroBooNE:2021ppm}   &&&& &       \\
& $\frac{d^2\sigma}{dp_\mu dp_p}$\cite{MicroBooNE:2020akw,MicroBooNE:2020fxd}    &&&& &       \\
&&&&& \\
NO$\nu$A & $\nu_{e}$\cite{NOvA:2022see} & & NC~\cite{NOvA:2019bdw} & -     & CH$_2$      & 2.0-2.0 \\
& $\frac{d^2\sigma}{dT_\mu d\cos{\theta_\mu}}$\cite{NOvA:2021eqi}    &&&& & \\
&&&&& \\
T2K & $\nu_{\mu}$\cite{T2K:2013nor,T2K:2018lnf,T2K:2014axs,T2K:2015cxp,T2K:2019dgm} & 1.26\cite{T2K:2014hih} & CC\cite{T2K:2021naz,T2K:2019yqu,T2K:2016cbz,T2K:2016soz} & & CH      & 0.6-0.6 \\
& $\nu_{e}$\cite{T2K:2014lbi,T2K:2015ydf,T2K:2020lrr} &&&& H$_2$O & \\
& $\bar{\nu}_{\mu}/\nu_\mu$\cite{T2K:2017bvo} &&&&Fe& \\
& $\frac{d^2\sigma}{dT_\mu d\cos{\theta_\mu}}$\cite{T2K:2020jav,T2K:2020sbd,T2K:2019ddy,T2K:2017qxv,T2K:2016jor}    &&&&  & \\
& $\sigma(E_\nu)$\cite{T2K:2015ujp}    &&&& & \\
& NC\cite{T2K:2014vog}    &&&& & \\
& $\frac{d\sigma}{d\delta p_T}\frac{d\sigma}{d\delta\alpha_{T}}$ \cite{T2K:2018rnz}    &&&& & \\
& O/C\cite{T2K:2020jav}    &&&& & \\
&&&&& \\

\end{tabular}
\end{table*} 

Table~~\ref{tab:xsection_results} provides a list of recent measurements, organized by experiment, 
their measurements for neutrino and anti-neutrino inclusive and differential cross section ($Meas$), cross section with charged pion in final state ($\pi^{\pm}$) for both CC and NC reactions, cross section with neutral pion in final state ($\pi^{\circ}$) for both CC and NC reactions, nuclear targets and beam properties for both the neutrino and antineutrino data. In Table~\ref{tab:xsection_results}, $M_A$ refers to the axial mass that is a parameter in the dipole parametrization of the nucleon axial form factor:
\begin{align}
F_A(Q^2) = \frac{g_A}{\left[ 1 + (Q^2/M_A^2) \right]^2} \ .
\end{align}

The cross sections of inclusive neutrino and antineutrino reactions\textemdash involving the elementary processes $\nu_\mu \rightarrow \mu^- X$ 
and $\bar{\nu}_\mu \rightarrow \mu^+ X$, respectively\textemdash 
have been measured at different beam energies using a variety of target.

Pion-less, that is, quasi elastic scattering, is the dominant neutrino interaction in the low-energy regime of $E_\nu <$1~GeV, where most of the signal sample for current neutrino oscillation experiments is found. For this reason, the pion-less regime has been analyzed in detail and deeply scrutinized in recent years. In the past, quasi elastic scattering was associated with events in which only one lepton is detected in the final state, and a single nucleon is emitted 
by the target nucleus in the primary neutrino interaction. These interactions were very easy to identify in bubble chamber experiments,  
where light nuclear targets, such as hydrogen or deuterium, were used.
In the case of heavier targets, however, the picture complicates substantially. Nuclear effects impact the cross section, both in absolute normalization and overall shape (bin-to-bin) distortions, and the number of final state particles will change as well as the kinematics. The simple assumptions that just one particle and a single nucleon participate in the scattering is not sufficient anymore to describe the primary interaction. Due to intranuclear rescattering and long- and short-range correlations between the nucleons, additional particles can be emitted in the final state; therefore, the quasi-elastic process no longer involves the emission of a single lepton and a single nucleon only. For this reason recent heavy target experiment, like MicroBooNE, started to refer to the quasi elastic reaction as pion-less. Many of these measurements are reported in Tab.~\ref{tab:xsection_results} and nucleon-only cross sections as a function of observed final state particle kinematics~\cite{MINERvA:2018hqn,MINERvA:2018vjb,MINERvA:2019gsf,MINERvA:2018hba,MiniBooNE:2010bsu,MiniBooNE:2013qnd,T2K:2016jor,T2K:2017qxv,T2K:2019ddy,T2K:2020jav,T2K:2020sbd,MicroBooNE:2020akw,MicroBooNE:2020fxd}. Those measurements are difficult to cross-compare but could be less model-dependent, and provide more constraints on theory than historical cross section, if they were given as a function of $E_\nu$ and $Q^2$. Some work has been done in this respect in Ref.~\cite{Mahn:2018mai}. In the past quasi elastic double differential cross section measurements turned out to be larger than expected, and this finding generated a huge amount of theoretical and experimental work. 
In many instances, angular distributions and momentum distributions have been shown to be in disagreement with expectation~\cite{MicroBooNE:2018xad,MicroBooNE:2020akw,MicroBooNE:2020fxd} and once the predictions are refined and the nuclear model used to describe neutrino-nucleus scattering expanded, from just including single nucleon knockout, the data/Monte Carlo agreement greatly improves. The data/MC agreement is not yet at the level of few percent as required by the future precision oscillation experiments, but it is improving, and the accuracy of nuclear theory predictions are improving as well. Fundamental in this new development will be the use of past and current electron scattering data that will offer a framework where theory models could be tested with quasi-mono-energetic electrons and with different levels of final state interactions.

Neutrinos and anti-neutrinos can also interact in-elastically, and produce in their interactions a nucleon exited state, like a $\Delta$ and other baryons. This is the regime that current and future experiments will use to collect neutrino events. Baryons rapidly decay and generate one or more pions in final state and a nucleon. Modern measurements, also in regards to this channel, have been performed on a variety of targets, including argon. In pion production interactions, nuclear effects and final state interactions play an even bigger role than in quasi-elastic interactions. Many experiments have reported measurements without carefully accounting for nuclear effects and the large uncertainties that such measurements carry. The experiment that provided the most comprehensive cross section measurements has been MiniBooNE, that reported a total of 16 single and double differential cross sections with both final state muons and pions~\cite{MiniBooNE:2010eis,MiniBooNE:2009koj,MiniBooNE:2010cxl,MiniBooNE:2009dxl}. MINER$\nu$A has achieved similar collections of measurements as MiniBooNE but at higher neutrino energies~\cite{MINERvA:2016sfc,MINERvA:2019rhx,MINERvA:2019rhx,MINERvA:2017okh}. ArgoNeuT and MicroBooNE have been adding new information on argon. Also in this case, double differential cross section have been providing information with less dependence on theoretical models. 

Both MINER$\nu$A and MicroBooNE have been using tuning of neutrino interaction generator to present their data. Such tuning, while correctly including systematic errors due to nuclear models and effects and detector efficiency and background estimations, makes the comparison between experiments very difficult. The tuning method folds into cross section measurements fudge parameters to tune the response of neutrino interaction generators and makes the task of reaching a coherent and accurate description of neutrino-nucleus interactions very challenging. Also the interpretation of the tuning parameters is very complicated due to the fact that there are overlaps between the pion-less and pion interaction regions, where interference and double counting play a major role.

\begin{figure}[!ht]%
\centering
\includegraphics[width=0.40\textwidth]{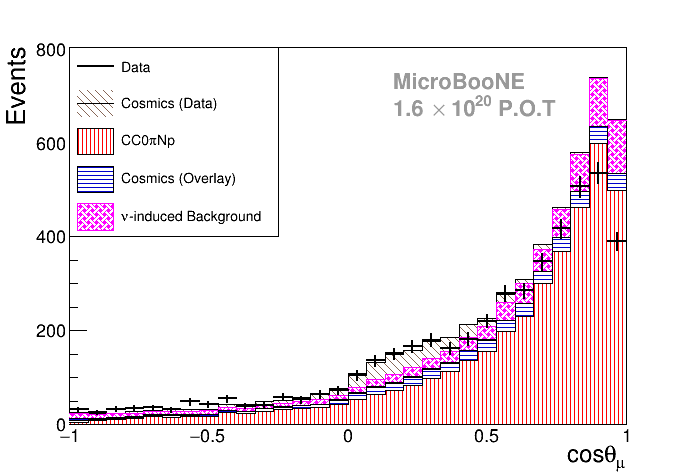}
\caption{MicroBooNE track angle distribution of the muon candidate in the CC-0$\pi$ sample and the leading proton candidate (cos$\theta$). MicroBooNE analysis is described in details in Ref.~\cite{MicroBooNE:2020akw}. Figure from Ref.~\cite{MicroBooNE:2020akw}.}
\label{fig:microboone_cc0pi}
\end{figure}

\begin{figure}[!ht]%
\centering
\includegraphics[width=0.45\textwidth]{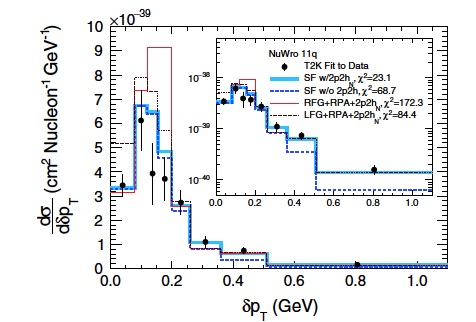}
\caption{The T2K extracted differential cross section as a function of the single transverse variables $\delta p_T$ compared to various MC included in the T2K analysis as described in Ref.~\cite{T2K:2018rnz}. Figure from Ref.~\cite{T2K:2018rnz}}
\label{fig:t2k_ccqe_mcs}
\end{figure}

The MicroBooNE~\cite{MicroBooNE:2020akw} experiment in the CC-0$\pi$ sample and the leading proton candidate (cos$\theta$) analysis noted a difference between the data and the MC predictions, see Fig.~\ref{fig:microboone_cc0pi}. The T2K~\cite{T2K:2018rnz} experiment also measured a difference between data and MC predictions in the extracted differential cross section as a function of the single transverse variables $\delta p_T$, see Fig.~\ref{fig:t2k_ccqe_mcs}. These differences could be easily attributed to the fact that the neutrino-nucleus interactions are not well modelled or that are not fully implemented in MC generators due to computational requirements..

\section{Theoretical models of the neutrino-nucleus interactions}
\label{nuA:th}

\begin{figure}[!h]%
\centering
\includegraphics[width=0.425\textwidth]{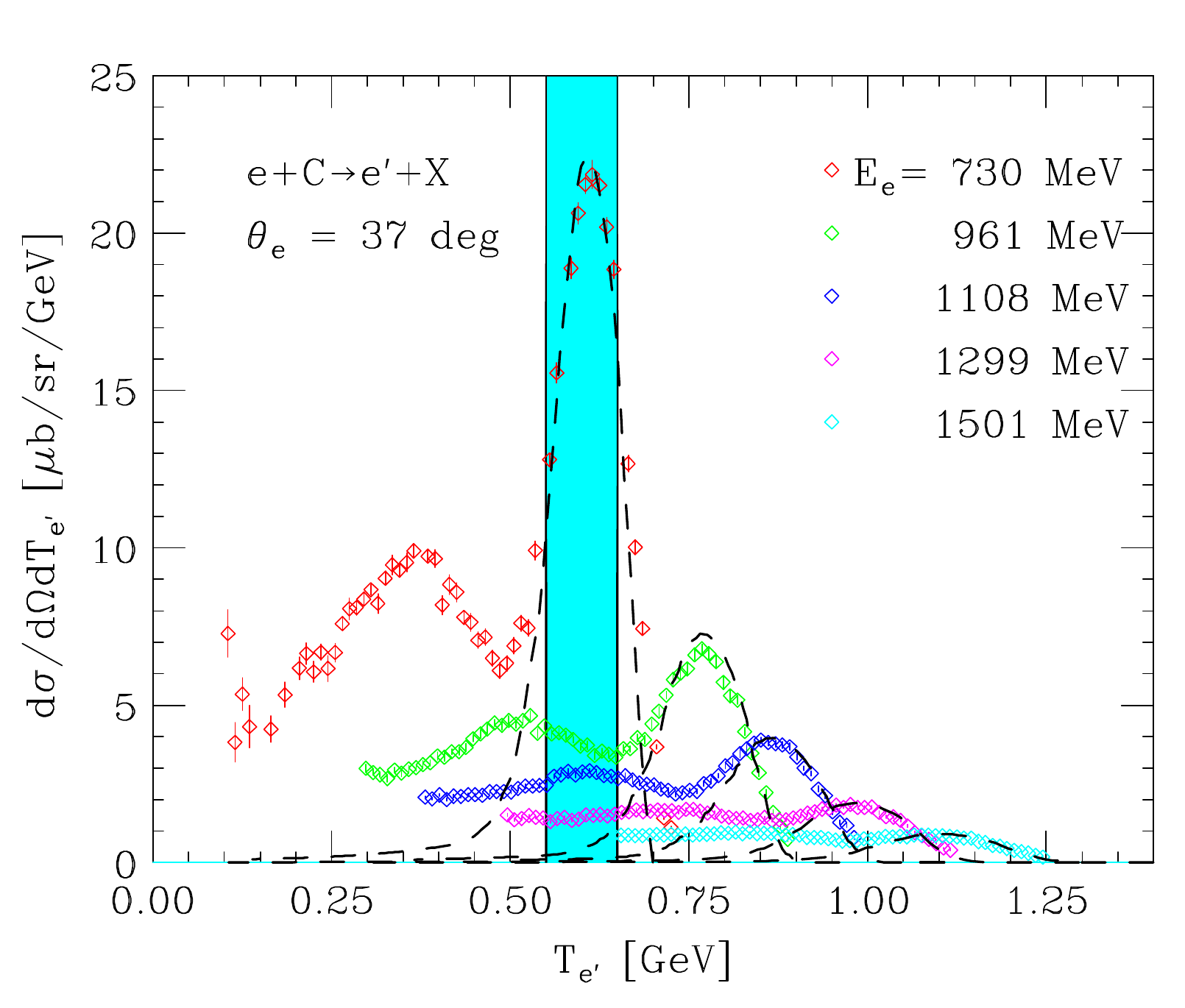}
\caption{Inclusive electron-carbon cross sections at electron scattering angle $\theta_e=$ 37 deg and beam energies ranging
between 0.730 and 1.501 GeV \cite{12C1,12C2}. The dashed lines represent the single nucleon 
knock out contribution, computed within the factorisation approach using the spectral function 
of Ref.~\cite{LDA}. From Ref.~\cite{Benhar:NUFACT11}.\label{e:flux}}
\end{figure}

The difficulties involved in the interpretation of the flux-integrated 
neutrino-nucleus cross section are illustrated 
in Fig.~\ref{e:flux}, in which the {\it electron}-carbon cross sections corresponding to 
the same scattering angle and different beam energies are displayed as a function 
of the energy of the scattered electron, $T_{e^\prime}$.
It clearly appears that, depending on beam energy, the cross section in the highlighted bin, corresponding to $0.55 \leq T_{e^\prime} \leq 0.65$ GeV, receives contributions 
from different reaction mechanisms. At $E_e = 730$ MeV quasi elastic single-nucleon 
knock out is largely dominant, while as the beam energy increases different processes, 
such as resonance production and deep-inelastic scattering, also provide sizable contributions.

Figure~\ref{e:flux} clearly implies that the description of the signal detected by 
neutrino experiments must be based on a model capable to take into account all 
relevant reaction mechanisms within a unified and consistent framework.

\subsection{The lepton-nucleus cross section}

The differential cross section of the lepton-nucleus scattering process
\begin{equation}
\label{reaction}
\ell + A \to \ell^\prime + X \   ,
\end{equation}
where $A$ and $X$ denote the target nucleus of mass number $A$
and the hadronic final state, respectively,
can be schematically written in the form
\begin{equation}
\label{dsigmaA:1}
{d \sigma}_A  \propto  L_{\alpha \beta} W_A^{\alpha \beta} \ ,
\end{equation}
where $L_{\alpha \beta}$ is fully specified by the lepton kinematic variables, 
while the tensor describing the target response
\begin{align}
\label{dsigmaA:2}
W_A^{\alpha \beta} & = \sum_X \left[ \langle 0 \vert {J^\alpha}^\dagger \vert X \rangle 
\langle X \vert {J^\beta} \vert 0 \rangle + {\rm h.c.} \right] \\ 
\nonumber
& \times \delta^{(4)}(P_0 + q - P_X)  \ ,
\end{align}
contains all the information on nuclear structure and dynamics. The above equation shows that
the description of the nuclear response involves the target initial and final states, carrying four-momenta $P_0$ and $P_X$, as well as the nuclear current operator
\begin{align}
\label{nuclear:current}
 {J^\alpha_A} = \sum_i {j^\alpha_i} + \sum_{j>i} {j^\alpha_{ij}} 
 = J_1^\alpha + J_2^\alpha ,
\end{align}
comprising both one- and two-nucleon terms.
The sum in Eq.~(\ref{dsigmaA:2}) includes contributions from all possible final states, excited through different reaction mechanisms whose relative importance 
depends on kinematics.

In neutrino-nucleus scattering, Charged Current Quasi Elastic (CCQE) scattering off an individual nucleon, that is, the process corresponding to the final state
\begin{equation}
\label{1p1h}
\vert X \rangle = \vert p, (A-1)  \rangle \ ,
\end{equation}
is the dominant mechanism in the kinematic region relevant to the analysis of, e.g.,  the MiniBooNE data, collected using
a neutrino flux of mean energy $\langle E_\nu \rangle~=~880~MeV$.

From the observational point of view, CCQE processes
are specified by the absence of pions in the final state. 
They are collectively referred to as $0\pi$ events, and are fully determined
by the measured kinetic energy and emission angle of the muon, with the knocked out proton and the recoiling nucleus being undetected.
The spectator $(A-1)$-nucleon system is left either in a bound state or in a state comprising
a nucleon excited to the continuum.

For example, for a carbon target the states of the recoiling
system are $\vert \isotope[11][]{C^*} \rangle$,  $\vert  p , \isotope[10][]{B^*}   \rangle$ or $\vert  n,  \isotope[10][]{C^*}  \rangle$, where the asterisk indicates that
the nucleus can be in its ground state or in any excited bound  states. The corresponding $A$-nucleon final states are
\begin{equation}
\label{1p1h_C}
\vert X \rangle = \vert p,  \isotope[11][]{C^*}  \rangle  \ ,
\end{equation}
or
\begin{align}
\label{2p2h_C}
\vert X \rangle = \vert p p,  \isotope[10][]{B^*}  \rangle \ \ , 
\ \  \vert p n,  \isotope[10][]{C^*}  \rangle \ ,
\end{align}
The states appearing in the right-hand side of Eqs.~(\ref{1p1h_C}) 
and  (\ref{2p2h_C}) are referred to as
one-particle\textendash one-hole (1p1h) and two-particle\textendash two-hole (2p2h), respectively.

The occurrence of 2p2h final states in processes in which the beam particle couples to an individual nucleon originates
from nucleon-nucleon correlations in the target ground state or final state interactions (FSI) between the struck particle and the spectator nucleons.
These mechanisms are not taken into account by models based on the independent particle picture of the nucleus, such as the RFGM, according to
which single-nucleon knock out  can only lead to transitions to 1p1h final states. On the other hand, transitions to 2p2h states are always allowed in processes driven by two-nucleon
meson-exchange currents (MEC), see Eq.(\ref{nuclear:current}), such as those
in which the beam particle couples to a $\pi$-meson exchanged between two interacting nucleons.
A detailed discussion of the contributions of 1p1h and 2p2h final states to the nuclear response can be found in Refs.~\cite{Martini2010,nieves,Benhar2015a,NoemiPRL,RBBDL}.
Obviously, the amplitudes of processes involving one- and two-nucleon currents and the same 2p2h final state
contribute to the nuclear cross section both individually and through interference.

More complex final states, that can be written as a superposition of 1p1h states according to
\begin{equation}
\label{RPA_C}
\vert X \rangle = \sum_n C_n \vert p_n h_n \rangle  \ ,
\end{equation}
appear in processes in which the momentum transfer is shared between many nucleons.
The contribution of these processes is often described in perturbation theory within the Random Phase Approximation  (RPA)\textemdash
which amounts to taking into account the so-called ring diagrams to all orders\textemdash using phenomenological effective interactions to describe nuclear dynamics~ \cite{Alberico,Gil1997543}.
On the basis of very general quantum-mechanical considerations,  long-range correlations associated with the final states of Eq.~(\ref{RPA_C}) are expected to become
important in the kinematic region in which the space resolution of the beam particle is much larger than the average
nucleon-nucleon distance in the nuclear target, $d$, i.e. for typical momentum transfers $\vert{\bf q}\vert \ll  \pi/d \sim 400  \ {\rm MeV}$.

\subsection{Impulse approximation, factorisation and spectral function}

The application of the formalism outlined in the previous section in the kinematic regime relevant to accelerator-based 
neutrino experiments requires a relativistic description of both the final state $ \vert X \rangle $, comprising a high-momentum 
nucleon, and the current operator, which depends explicitly on momentum transfer. On the other hand, the initial state of
the target can be safely treated using non relativistic many-body theory.

 The IA, extensively employed to analyze electron-nucleus scattering data~\cite{Benhar:2006wy}, is based on the tenet that,
at momentum transfer ${\vert {\bf q} \vert}$ such that ${\vert {\bf q} \vert} \gg  \pi/d$,  
neutrino-nucleus scattering reduces to the incoherent sum of scattering processes involving individual nucleons. In addition, final state interactions (FSI) between the outgoing hadrons and the spectator nucleons are expected to be small, and tractable as corrections.

Within the IA picture, the nuclear current of Eq.\eqref{nuclear:current} reduces to the sum of one-body operators,
while the final state simplifies to the direct product of the hadronic state produced at the interaction vertex 
with momentum ${\bf p}$, and the state describing the $(A-1)$-nucleon recoiling system, carrying momentum ${\bf p}-{\bf q}$
\begin{align}
\label{fact:1}
\vert X \rangle = \vert {\bf p} \rangle \otimes \vert n_{(A-1)} \rangle .
\end{align}
As a consequence, the nuclear transition matrix element can be written in the {\it factorised} form
\begin{align}
\label{ampl:1}
\langle X \vert J^\alpha_1 \vert 0 \rangle = \sum_i \int d^3 k \ M_n({\bf k}) 
\langle {\bf k} + {\bf q} \vert j_i^\alpha \vert {\bf k} \rangle \ , 
\end{align}
with
\begin{align}
\label{amplitude:1}
M_n({\bf k}) = \big\{ \langle n_{(A-1)} \vert \otimes \langle {\bf k} \vert  \vert 0 \rangle  \big\}, 
\end{align}
where ${\bf k} = {\bf p}-{\bf q}$ is the momentum of the struck nucleon. 
The resulting nuclear cross section reduces to
\begin{align}
    d\sigma_A =  \int d^3k dE \ d\sigma_{N} \ P({\bf k},E) \ .
\label{fact:xsec}
\end{align}
In the above equation, $d\sigma_{N}$ is the differential cross section of the 
elementary scattering process involving a {\it moving bound nucleon}, while the spectral function 
\begin{align}
P({\bf k},E) = \sum_n \vert M_n({\bf k}) \vert^2 \ \delta(E+E_0-E_n) \ , 
\end{align}

with $E_0$ and $E_n$ being, respectively, the energies of the states $\vert 0 \rangle$ and  $\vert n_{(A-1)} \rangle$, 
describes the probability of removing a nucleon of momentum ${\bf k}$ from the nuclear ground state leaving the residual system with excitation energy $E$. The effect of binding can be taken into account following a procedure originally 
proposed in Ref.~\cite{forest83} and widely employed in the analysis of electon scattering data
~\cite{Benhar:2006wy}.

\begin{figure}[!h]%
\centering
\includegraphics[width=0.40\textwidth]{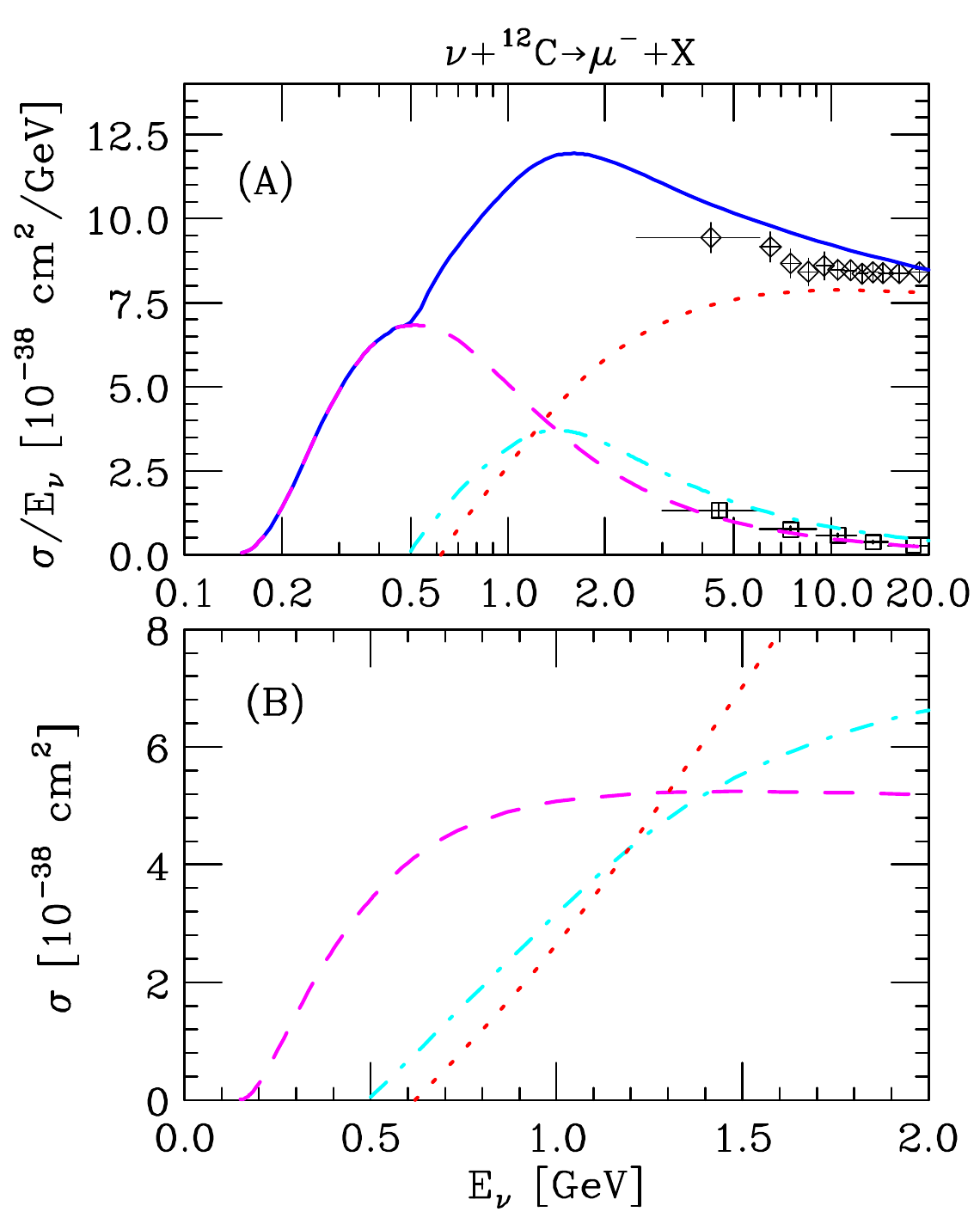}
\caption{Total cross section of the reaction $\nu_\mu + {^{12}C} \to \mu^- + X$ as a function of neutrino energy. 
The dashed, dot-dash and dotted lines of panel (B)
represent the contributions of CCQE, RES and DIS processes. Panel (A) shows the $E_\nu$-dependence of the ratio $\sigma/E_\nu$.
The meaning of the dashed, dot-dash and dotted lines is the same as in panel (B). The full line corresponds to the sum of the three contributions.
Diamonds and squares represent the data of Refs.~\cite{NOMAD:2007krq} and \cite{NOMAD:2009qmu}, respectively.
From Ref.~\cite{Vagnoni:2017hll}.}\label{erica}.
\end{figure}

Note that the spectral function describes intrinsic properties of the target, independent of the nature of the interaction vertex. Therefore, depending on the form of $d\sigma_{N}$, Eq.\eqref{fact:xsec} can be used to obtain 
the electron- and neutrino-nucleus cross sections in any reaction channels. 
For example, in CCQE neutrino interactions the nucleon cross section involves the vector and axial form factors of the nucleon, whereas the resonance production (RES) and 
deep-inelastic scattering (DIS) cross sections are written in terms of inelastic structure functions~\cite{BenharMeloni,Vagnoni:2017hll}.  

Figure~\ref{erica} shows the energy dependence of the results of a calculation of the total neutrino-carbon cross section, carried out using Eq.~\eqref{fact:xsec} with the spectral function of Ref.~\cite{LDA}, the vector form factors of Ref.~\cite{BBBA}, and the dipole parametrisation of the axial form factor~\cite{Vagnoni:2017hll}.
For comparison, the breakdown of the full result into QE, RES, and DIS contributions is also displayed. 

It is apparent that, while at beam energy $E_\nu \lesssim 0.8$ GeV QE interactions dominate,
the inelastic cross section rapidly increases with energy; at  $E_\nu \approx 1.3$~GeV, the contributions arising from the three active reaction channels turn out to be about the same.
For comparison, panel (B) also reports, as diamonds and squares, the total $\nu_\mu$-carbon cross section measured by the NOMAD collaboration \cite{NOMAD:2007krq}.
It turns out that, although the energy-dependence of the data at $E_\nu \gtrsim 10$ GeV is well reproduced  by the theoretical prediction of the DIS cross section, 
the results of the full calculation sizably exceed the data. In view of the fact that the 
QE cross section reported by the NOMAD Collaboration~\cite{NOMAD:2009qmu} turns out to be in close agreement with the theoretical results, 
this discrepancy is likely to be ascribed to double counting of the RES and DIS contributions, which are very hard to identify in a truly model independent fashion.

\begin{figure*}[!h]%
\centering
\includegraphics[width=0.75\textwidth]{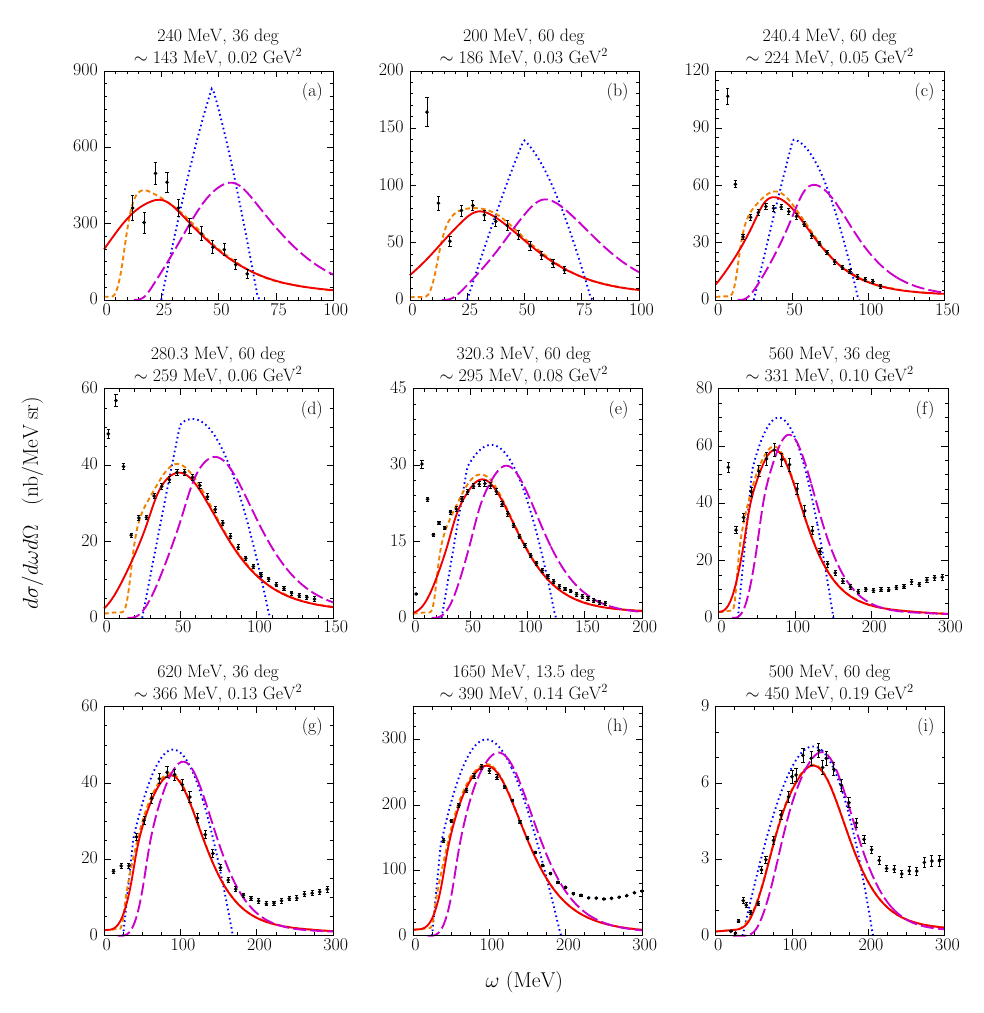}
\caption{Double differential electron-carbon cross sections in the QE channel, computed by the authors of Ref.~\cite{Ankowski2013}
within the spectral function approach, compared to the data of
Ref.~\cite{Barreau83,Baran88,Whitney74}. The solid lines correspond to the result of the full calculation, whereas the long-dashed lines have been obtained neglecting FSI.
The difference between the solid and short-dashed lines illustrates the effect of using alternative treatments of Pauli blocking. For comparison, the prediction of the RFGM are
also shown, by the dotted lines. The panels are labeled according to beam energy, scattering angle, and values of $|{\bf  q}|$ and $Q^2$ at the quasielastic peak, corresponding to $ \omega = Q^2/2m $. Adapted from Ref.~\cite{Ankowski2013}.}\label{artur}
\end{figure*}

\subsection{Corrections to the impulse approximation}

Corrections to the IA, arising primarily from FSI and MEC, can be consistently included in the factorisation scheme.

\subsubsection{Final State Interactions}

In QE scattering, the occurrence of final state interactions (FSI) between the outgoing nucleon and the recoiling system can be taken into account using  a conceptually straightforward extension of the formalism based on the nucleon spectral function~\cite{Benhar:FSI}.

Within this scheme, referred to as convolution approach~\cite{Sosnik91,Benhar91} the lepton-nucleus cross section is written in terms of the
IA result according to
\begin{align}
\label{convolution}
\frac{d \sigma_A}{d\Omega_{\ell^\prime} dE_{\ell^\prime}} = \int d\omega^\prime \ \Bigg( \frac{d \sigma_{A}}{d\Omega_{\ell^\prime} dE_{\ell^\prime}} \Bigg)_{IA} 
F_{\bf q}(\omega - \omega^\prime) \ ,
\end{align}
where $\omega$ is the energy transfer and 
\begin{align}
\label{def:ff}
F_{\bf q}(\omega) = \sqrt{T_A}\delta(\omega)  + (1 - \sqrt{T_A}) f_{\bf q}(\omega) \ .
\end{align}

The above definition of the {\it folding function}, embodying all FSI effects, comprises the nuclear transparency $T_A$\textemdash that can be extracted from the measured cross sections of the $(e,e^\prime p)$ reaction~\cite{Rohe05}\textemdash and a finite-width function $f_{\bf q}(\omega)$ sharply peaked at $\omega = 0$, whose calculation involves the nucleon density distributions and the nucleon-nucleon (NN) scattering amplitude at momentum $\vert {\bf q} \vert$~\cite{Benhar:FSI}. Note that both $T_A$ and $f_{\bf q}(\omega)$ are strongly affected by short-range correlations among nucleons in the target nucleus, because the repulsive core of the NN potential reduces the probability that the struck particle interact with one of the spectators
within a distance $\sim~1 \ {\rm fm}$ of the primary interaction vertex \cite{BenharCT}. In the absence of FSI, $T_A \to 1$, implying that
the residual nucleus is fully transparent to the outgoing struck nucleon, $f_{\bf q}(\omega) \to \delta(\omega)$, and the IA cross section is recovered.

The most prominent effects of FSI on the observed $\omega$-dependence of the nuclear cross section are: (i) a shift towards lower values of $\omega$, accounting for the energy needed to remove the struck nucleon
from the target nucleus, and (ii) a broadening of the cross section, arising from rescattering of the outgoing particle against the spectators, leading to a quenching of its peak and an enhancement of the tails~\cite{Ankowski2013}.

As an example, Fig.~\ref{artur} shows the results of a comprehensive analysis of the cross sections of the process
\begin{align}
e + \isotope[12][]{C} \to e^\prime + X \ , 
\label{eep:carbon}\end{align}
carried out by the authors of Ref.~\cite{Ankowski2013} using the factorisation formalism, 
 the carbon spectral function of Ref.~\cite{LDA}, and the nucleon form factors of Ref.~\cite{BBBA}. 
The impact of FSI corrections in a variety of kinematic conditions, corresponding to squared four-momentum transfer 
$0.02 \leq Q^2 \leq 0.49$ GeV$^2$, is illustrated by the differences between the dashed lines, representing the IA results, and the solid lines, corresponding to the full calculation. 
It is apparent that the approach based on factorisation and the spectral function 
formalism provides a remarkably accurate description of the full data set, spanning a broad kinematic range.  On the other hand, a comparison with the results obtained using 
the RFGM, displayed by the dotted lines, clearly indicates that, in most instances, this oversimplified model conspicuously fails to explain the data.

\subsubsection{Meson-Exchange Currents}
\label{section:MEC}

The description of reactions involving meson-exchange currents (MEC) requires an extension of the factorisation scheme, needed to 
include processes in which the beam particle interacts with {\it two} nucleons. In this case, the final state
appearing in Eq.~\eqref{reaction} is written in the form
\begin{align}
\vert X  \rangle = \vert {\bf p} , {\bf p}^\prime \rangle \otimes \vert n_{(A-2)} \rangle \ ,   
\end{align}
to be compared to Eq.~\eqref{fact:1}, and the transition amplitude 
reduces to
\begin{align}
\label{ampl:2}
& \langle X \vert J^\alpha_2 \vert 0 \rangle = \int d^3k d^3 k^\prime {\mathcal M}_n({\bf k},{\bf k}^\prime) \\
 \nonumber
 & \ \ \ \ \ \ \ \times \langle {\bf p} {\bf p}^\prime \vert j_{ij}^\mu \vert {\bf k} {\bf k}^\prime \rangle
 \ \delta({\bf k} + {\bf k}^\prime + {\bf q} - {\bf p} - {\bf p}^\prime) \ .
\end{align}
In the above equation the nuclear amplitude ${\mathcal M}_n({\bf k},{\bf k}^\prime)$\textemdash analogue to $M_n({\bf k})$ of Eq.~\eqref{amplitude:1}\textemdash involves the overlap between the target ground state
and the state of the recoiling $(A-2)$-particle spectator system, whose properties can be reliably calculated in non relativistic 
approximation~\cite{spec2}. The two-nucleon matrix element, on the other hand, involves the fully relativistic expression of the current and free-nucleon states. 

The {\it extended} factorisation scheme allows to take into account in a fully consistent fashion 
the contributions of one- and two-nucleon currents, interference terms included, as well as the 
occurrence of inelastic processes. 

\begin{figure}[!h]%
\centering
\includegraphics[width=0.35\textwidth]{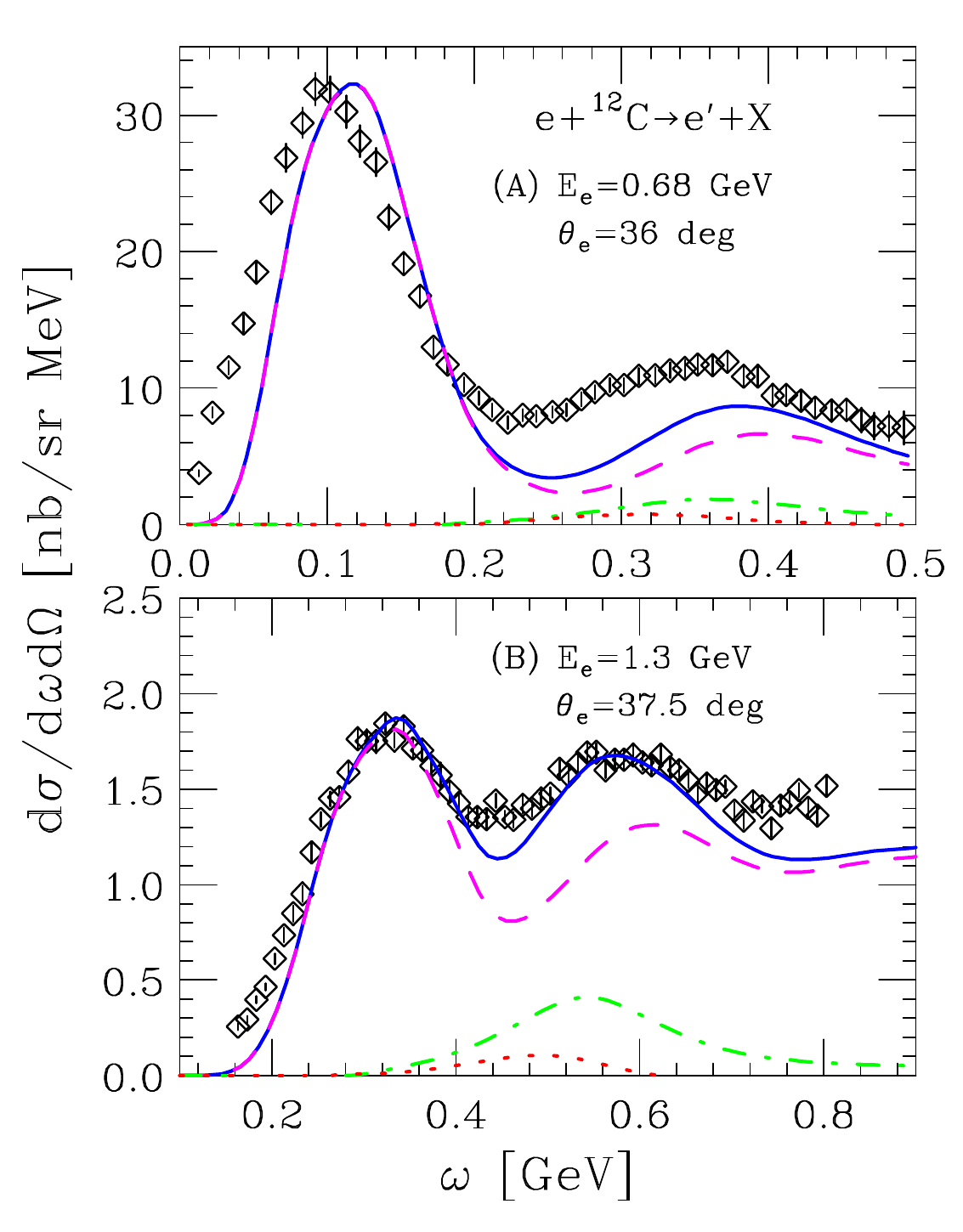}
\caption{(A) Double differential cross section of the process $e+{^{12}C}\to e^\prime +X$ at beam energy $E_e=680 $ MeV and scattering angle $\theta_e= 37.5$ deg. The solid line shows the results of calculations including MEC, while the dashed line has been obtained including only the one-body current. The effects of FSI corrections are not shown.
The contributions arising from the two-nucleon current are illustrated by the dot-dash and dotted lines, corresponding to the pure two-body current transition probability and
the interference term, respectively. The experimental data are taken from Ref.~\cite{Barreau83}; (B) same as (A) but for $E_e= 1300$ MeV and $\theta_e= 37.5$ deg. The experimental data are taken from Ref.~\cite{12C2}.
From Ref~\cite{NoemiPRL}.}\label{roccoPRL1} 
\end{figure}
\begin{figure}[!h]%
\centering
\includegraphics[width=0.35\textwidth]{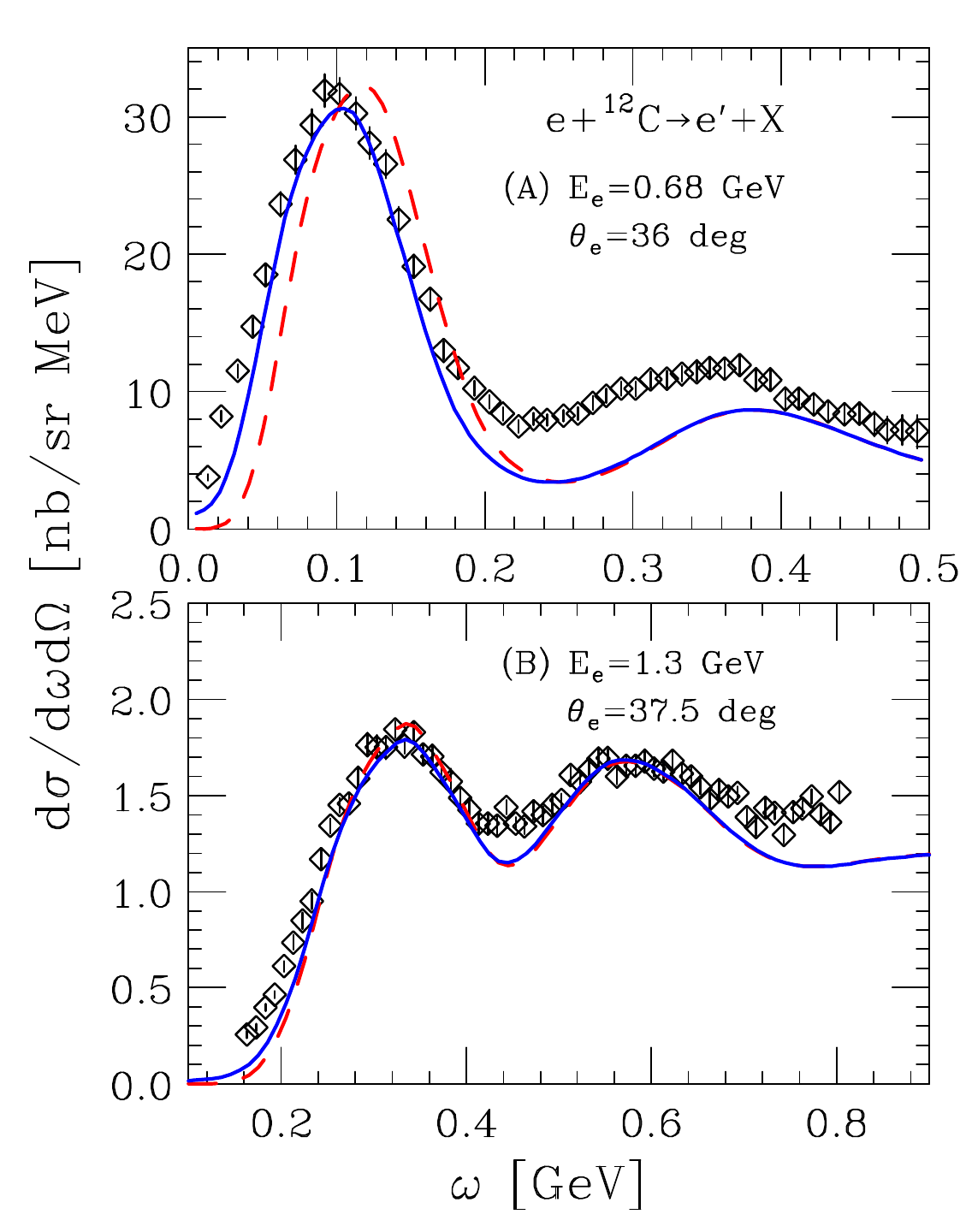}
\caption{(A): double differential electron-carbon cross section at beam energy $E_e=680 $ MeV and scattering angle $\theta_e= 36$ deg. The dashed line corresponds to the result obtained including MEC but neglecting FSI, while the solid line represent the results of the full calculation, including the combined effect of MEC and FSI. The experimental data are taken from Ref.~\cite{Barreau83}. (B): same as (A) but for $E_e= 1300$ MeV and $\theta_e= 37.5$ deg. The experimental data are taken from Ref.~\cite{12C2}. From Ref~\cite{NoemiPRL}.
}\label{roccoPRL2}
\end{figure}

As an example, the results of the analysis of electron-carbon scattering data carried out by the authors of Ref~\cite{NoemiPRL} 
including both FSI and MEC corrections to the IA cross section, 
are shown in Figs.~\ref{roccoPRL1} and \ref{roccoPRL2}. 

Figure~\ref{roccoPRL1} illustrates the contribution of MEC alone, while 
the combined effect of MEC and FSI can be appreciated from the cross sections displayed in 
Fig.~\ref{roccoPRL2}.
A comparison with the results of the full theoretical calculations, represented by the solid lines of Fig.~\ref{roccoPRL2},  
shows that the data are reproduced with remarkable accuracy in the region of the quasi elastic peak, 
where single nucleon knock out dominate, as well as in the dip region between the quasi elastic and the 
$\Delta$ production peak, in which the contribution of MEC turns out to be significant. As argued by the authors of Ref.~\cite{BFNSS,BenharMeloni}, the discrepancy 
between theory and experiment in the resonance production region is likely to originate from the uncertainty
associated with the description of the neutron structure functions.

\section{The issue of flux average}
\label{flux:average}

\begin{figure*}[!h]%
\centering
\includegraphics[width=0.400\textwidth]{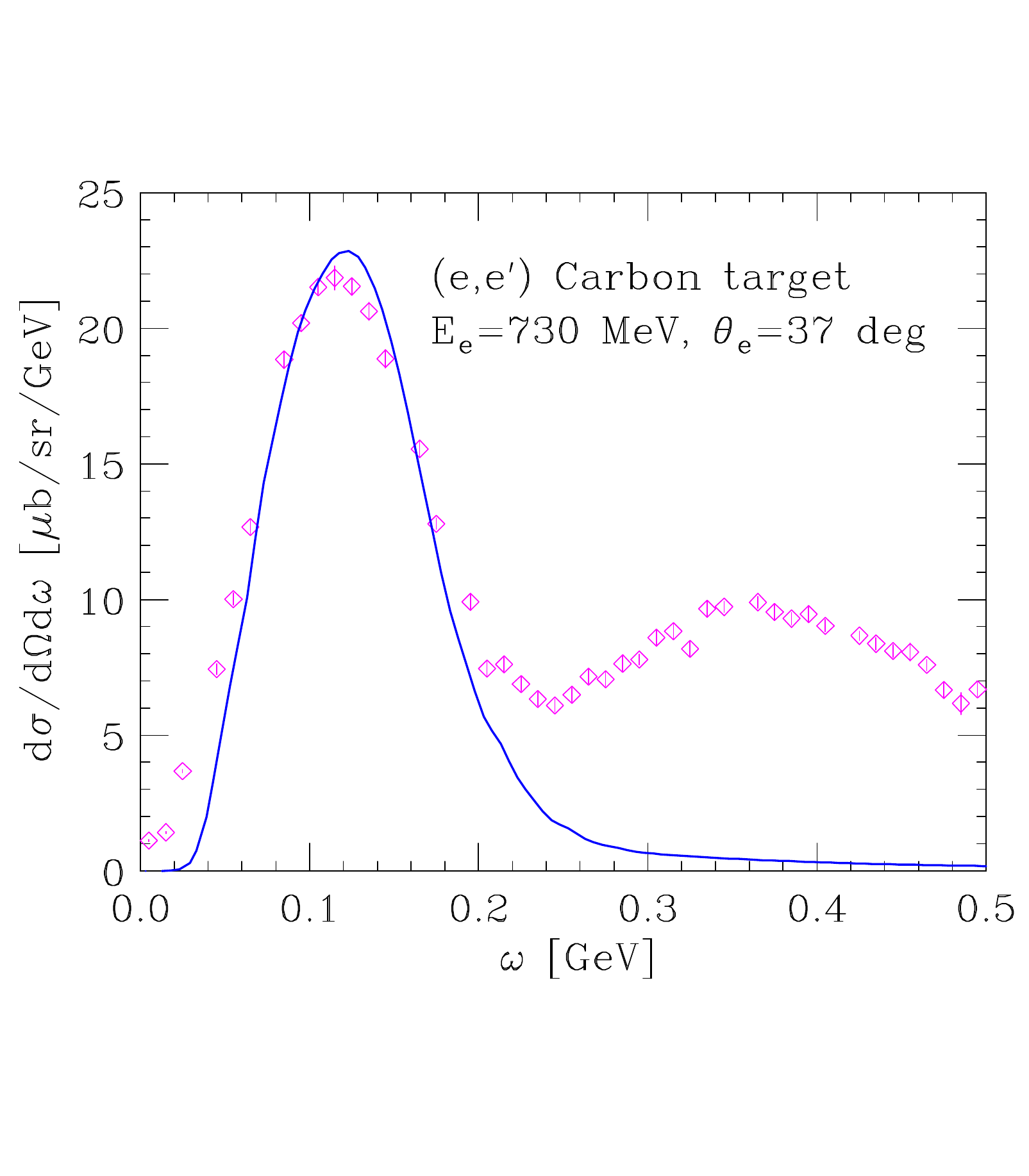}
\includegraphics[width=0.400\textwidth]{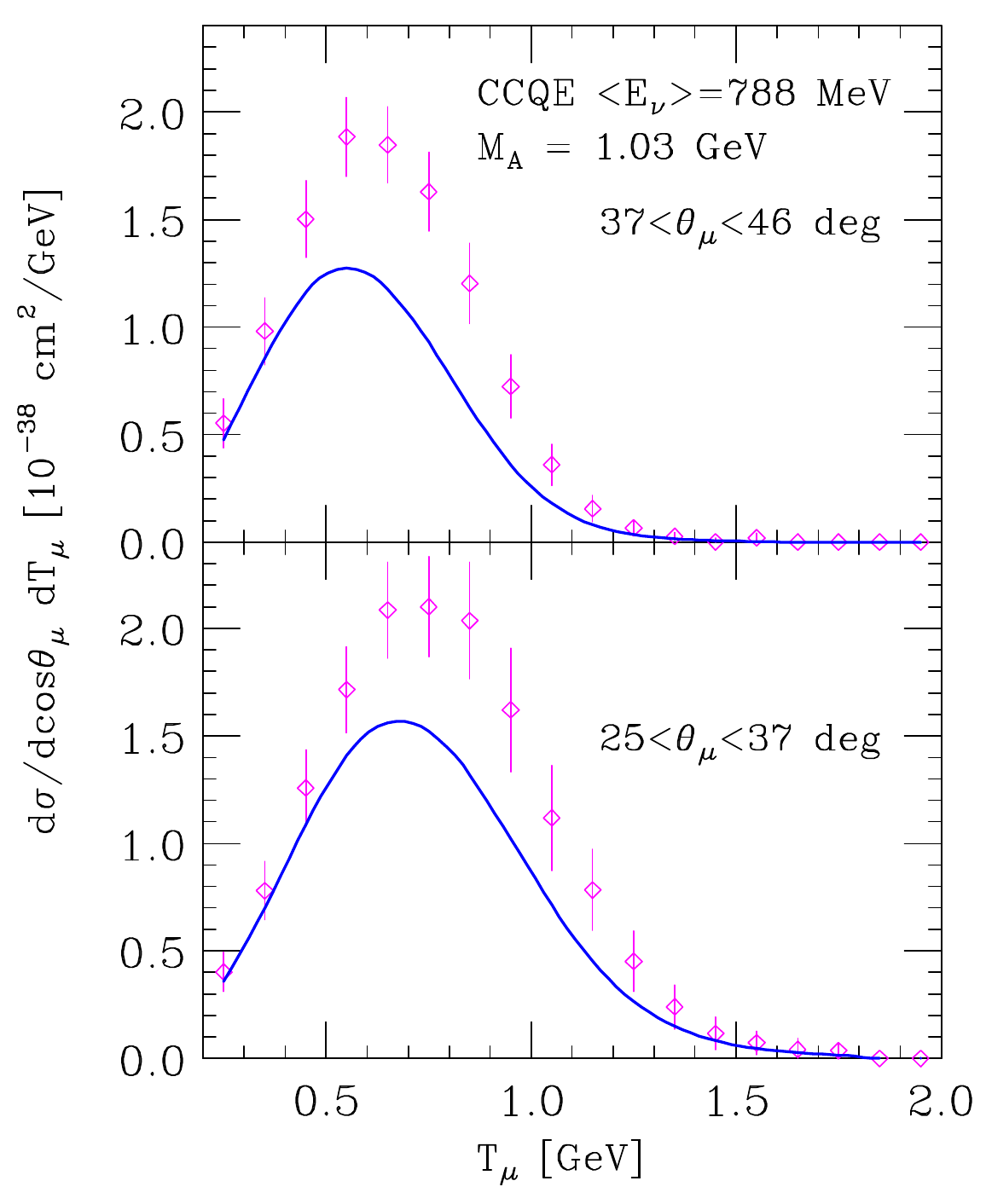}
\caption{Left: Inclusive electron-carbon cross section at beam energy $E_e=$~730~MeV and electron scattering
angle $\theta_e=37^\circ$, plotted as a function of energy loss $\omega$ \cite{Benhar:2010nx}. The data are from
Ref.~\cite{OConnell:1987kww}. Right: Flux averaged double differential CCQE cross section measured by the MiniBooNE collaboration
\cite{MiniBooNE:2009koj},  shown as a function of kinetic energy of the outgoing muon. The upper and lower panels 
correspond to
to different values of the muon scattering angle. Theoretical results have been obtained using the same spectral
functions and  vector form factors employed in the calculation of the electron scattering cross section of the
left panel and a dipole parametrisaition of the axial form factor with $M_A=1.03$ MeV. Adapted from Ref.~\cite{Benhar:2010nx}.}\label{e_nu:compare}
\end{figure*}

The difficulties implied in the theoretical description of the flux-averaged cross section 
are clearly illustrated in Fig.~\ref{e_nu:compare}. The left panel shows that the theoretical 
approach based on factorisation and the spectral function formalism provides a remarkably 
good description of electron scattering data in the quasi elastic channel\footnote{The capability to explain the data of Ref.~\cite{OConnell:1987kww} using the spectral function formalism  has been confirmed by the results of the improved and comprehensive analysis  performed by Ankowski {\it et al.}~\cite{Ankowski2013}.}; 
the results displayed in the the right panel, on the other hand, show that the 
same same approach fails to explain the flux-averaged 
neutrino cross section in seemingly comparable kinematic conditions. 

Unlike the electron scattering cross section of the left panel, the flux-averaged CCQE neutrino cross section
at fixed energy and scattering angle of the outgoing lepton collects the contributions 
of different kinematic regions, corresponding to a broad range of energies of the beam particles, in which different reaction mechanisms dominate. As a consequence, it cannot be described according to the scheme successfully applied 
to electron scattering, which is based on the tenet that the energy transfer to the 
target is fully determined by the measured kinematic variables.

As argued by the authors of Refs.~\cite{Benhar:Neutrino2010,Benhar:2010nx}, a meaningful description of the
flux-averaged cross section requires the adoption of a {\em new paradigm}, in which all relevant reaction mechanisms are accurately taken into account within a unified and consistent framework.

In the CCQE channel\textemdash which, from the observational point of view, is
characterised by the absence of $\pi$-mesons in the final state\textemdash 
single nucleon knock-out is dominant, and  the nuclear cross section 
is largely determined by the target spectral function and the nucleon form factors. 

The proton $(p)$ and neutron $(n)$ vector form factors, $F_1^{p,n}(Q^2)$ and
$F_2^{p,n}(Q^2)$, have been precisely measured
by electron-proton and electron-deuteron scattering experiments up to large values of $Q^2$;
see, e.g., Ref.~\cite{Perdrisat:2006hj}. 

The world average of the measured values of the axial mass, mainly
obtained from low statistics experiments carried out using deuterium targets, turns out
to be  $M_A~=~1.03~\pm~0.02$~GeV \cite{Bernard2002,Bodek:2007ym,NOMAD:2009qmu},
while the analyses performed by the K2K~\cite{K2K:2006odf} and MiniBooNE~\cite{MiniBooNE:2007iti}
collaborations using oxygen and carbon targets, respectively, yield
$M_A = 1.2$ and 1.35 GeV, respectively. The authors of Ref.~\cite{MiniBooNE:2007iti} suggested that a large value of $M_A$ should be interpreted as an {\it effective} axial mass, meant to parametrise nuclear effects not taken into account in the RFGM employed in data analysis. 

If the RFGM is replaced by a more realistic description of the carbon ground state, comprising
the admixture of 2p2h excitations, the normalisation of single nucleon states is sizably reduced from the 
unit value predicted by the mean-field approximation. As a consequence, the cross section obtained by the authors of Ref.~\cite{Benhar:2010nx} from state-of-the-art spectral function model, turns out to be lower that that predicted by the RFGM, and to bring theory into agreement with the MiniBooNE data the value of the axial mass must be increased to $M_A = 1.6$ GeV.

\begin{figure}[!h]%
\centering
\includegraphics[width=0.4500\textwidth]{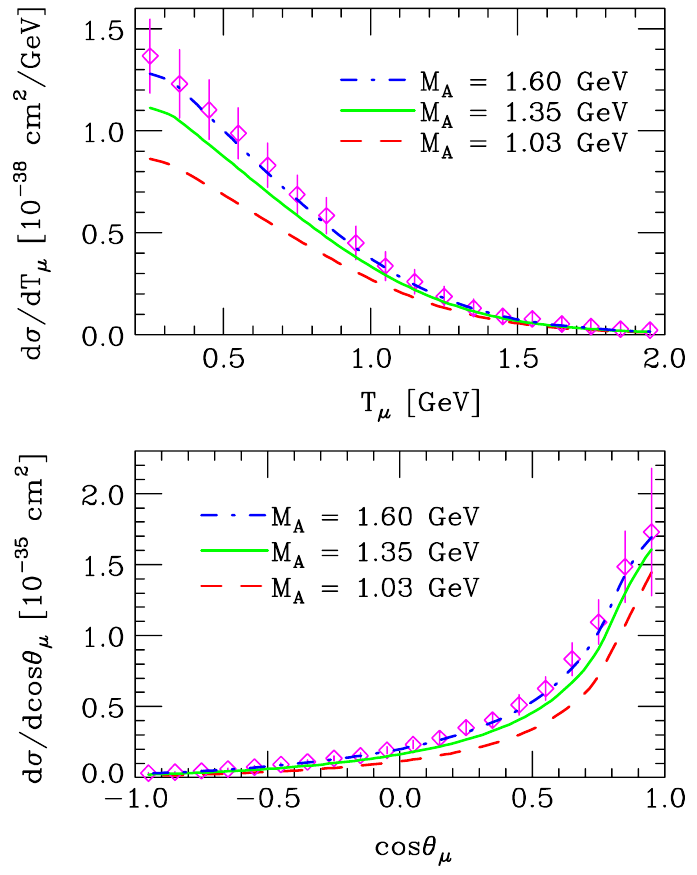}
\caption{Upper panel: flux-averaged muon kinetic energy spectrum 
of the  $\nu_\mu + \isotope[12][]{C} \to \mu^- + X$ reaction in the CCQE channel.
The dot-dash, solid and dashed lines have been obtained setting the value of the axial
mass to $M_A = 1.6, \ 1.35 \ {\rm and} \  1.03$ GeV, respectively.
The data are taken from Ref.\cite{MiniBooNE:2007iti}. Bottom panel: same as the upper panel, but for
the flux-averaged muon angular distribution. From Ref.~\cite{Benhar:2010nx}.}\label{distributions}
\end{figure}

This feature is illustrated in Fig.~\ref{distributions}, showing a comparison between the 
distributions of the MiniBooNE data as a function of kinetic energy (upper panel) and emission angle (lower panel) of the emitted muon and the corresponding theoretical results obtained
by the authors of Ref.~\cite{Benhar:2010nx} setting he value of the axial mass to 
1.03, 1.35, and 1.60 MeV. The results of the state-of-the-art study of Ankowski~\cite{PhysRevD.92.013007} show that, for $M_A~=~1.15$ GeV, the theoretical framework validated by the analysis of electron-carbon data of Ref.~\cite{Ankowski2013} provides an accurate account of the $Q^2$-distributions of the neutrino and antineutrino-carbon events collected by the MINER$\nu$A Collaboration~\cite{MINERvA:2013bcy,MINERvA:2013kdn}.

The difficulty to explain the MiniBooNE data with the single-nucleon knock out 
reaction triggered the development of theoretical models including the contributions of 
more complex reaction mechanisms. A prominent role in this context is played by 
processes involving 2p2h final states. 

Nuclear final states comprising two nucleons excited to continuum states above the Fermi surface
appear in the aftermath of {\it both} one- and two-nucleon knock out processes. 

In the case of single-nucleon knock out, the driving mechanisms are
NN corrrelations in the target initial state, or collisions involving the outgoing nucleon and the spectators. Note that, being obtained using a realistic spectral function, the flux-integrated cross sections shown in the right-hand panel of Fig.~\ref{e_nu:compare}  
includes the contribution of 2p2h final state excited by ground-state correlations.

Two-nucleon MEC, contribute to the cross sections through processes 
in which momentum and energy transfer are shared between two nucleons. As a consequence, the description 
of these reactions involves the two-nucleon spectral function, and is often obtained from the 
RFGM. 

\begin{figure}[!h]
\vspace*{-.05in}
\begin{center}
\includegraphics[scale=0.6]{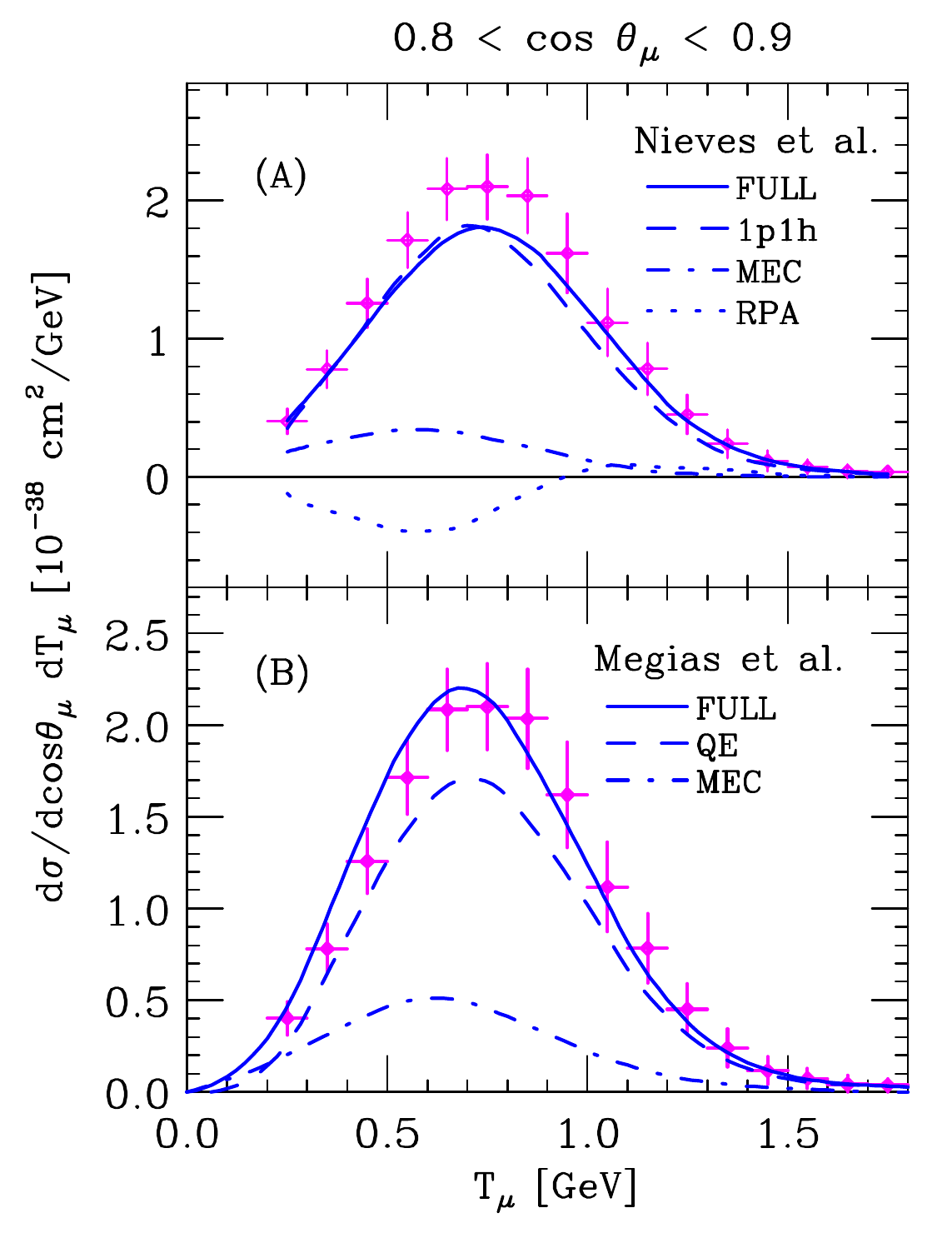}
\end{center}
\caption{Comparison between the double differential $\nu_\mu$-carbon cross section in the CCQE channel measured by the
MiniBooNE Collaboration~\cite{MiniBooNE:2007iti} and the
results obtained from the models of Nieves {\em et al.}~\cite{nieves} (A), and Megias {\em et al.}~\cite{SuSa}  (B). The solid lines
correspond to the full calculations. The meaning of the dashed, dot-dash and dotted lines is explained
in the text. Adapted from Refs.~\cite{nieves} and \cite{SuSa}.}
\label{degeneracy}
\end{figure}

The role played by the reaction mechanisms taken into account by two different models of neutrino-nucleus interactions is 
illustrated in Fig.~\ref{degeneracy}, showing a comparison between the flux-integrated double-differential CCQE cross section 
measured by the MiniBooNE Collaboration~\cite{MiniBooNE:2007iti} and the theoretical results of 
Nieves {\it et al.}~\cite{nieves} [Fig.~\ref{degeneracy}, panel (A)], and Megias {\it et al.}~\cite{SuSa} [Fig.~\ref{degeneracy}, panel (B)].

Within the approach of Ref.~\cite{nieves}, interactions leading to the appearance of 
1p1h final states are described 
within the {\it local} RFGM,  in which the Fermi momentum 
is obtained from the measured charge-density distribution of the target nucleus through the relation
$k_F(r) = 3 \pi^2 \rho^{1/3}_{\rm ch}(r)$. The 1p1h contribution to the CCQE cross section is supplemented with those arising from processes involving both MEC and long-range RPA correlations, obtained from the model originally proposed in Ref.~\cite{RPA2}.

The hybrid model of Ref.~\cite{SuSa} combines the results of the phenomenological scaling analysis of electron scattering
data \cite{Donnelly_superscaling} with a theoretical calculation of MEC contributions, carried out within the RFGM. Note that this scheme 
in inherently inconsistent, and does not allow to take into account the interference between one- and
two-body current contributions.

Overall, Fig.~\ref{degeneracy} shows that, up to a 10\% normalization uncertainty~\cite{nieves}, the two models provide comparable
descriptions of the data. However, their predictions are obtained from the combination of {\it different} reaction mechanisms.

The results reported in panel (A) indicate that, according to the model of Ref.~\cite{nieves}, the calculation including 1p1h final states only (dashed line labelled 1p1h) yields
a good approximation to the full result, represented by the solid line. The corrections arising from MEC (dot-dash line) 
and long-range correlations (dotted line labelled RPA) turn out to largely cancel one another. On the other hand, panel (B) suggests that, once
single-nucleon knock out processes (dashed line labelled QE) and MEC (dot-dash line) are taken into account, the addition of long-range correlations\textemdash whose contribution does not exhibit scaling\textemdash is {\it not} needed to explain the data.

The picture emerging from Fig.~\ref{degeneracy} calls for a deeper scrutiny. First and foremost, 
it is important to realize that studying the role of mechanisms other than
the excitation of 1p1h final states is only relevant to the extent to which the 1p1h sector, providing the dominant contribution to the cross section, is understood at fully quantitative level. In this context, the results shown in Fig.~\ref{degeneracy} appear to be somewhat misleading.

Detailed information on single nucleon knock out processes leading to the excitation of 1p1h final states has been obtained from experimental studies of the {\it exclusive} electron scattering process
\begin{align}
e + A \to e^\prime + p + (A-1)_{\rm B} \ ,
\label{eep}
\end{align}
in which the scattered electron and the outgoing proton
are detected in coincidence, and the recoiling nucleus is left in a bound state.
Up to calculable FSI corrections, the $A(e,e^\prime p)$ cross section provides access to the 
target spectral function, describing the energy and momentum distribution of the struck nucleon. 
The peaks corresponding to single nucleon emission can be identified in the missing energy spectra 
obtained from the analysis of $(e,e^\prime p)$ data, and the momentum-space wave functions of 
the single-nucleon states are directly related to the 
missing momentum distributions. As pointed out above, the spectroscopic factors, yielding the normalisation 
of single-nucleon states, sizably deviate from the unit value predicted by mean-field models.  
A fraction of the normalisation of the ground-state wave function is, in fact, provided by the 2p2h component,  
arising from NN correlations.

\begin{figure}[h!]
\vspace*{-.05in}
\begin{center}
\includegraphics[scale=0.425]{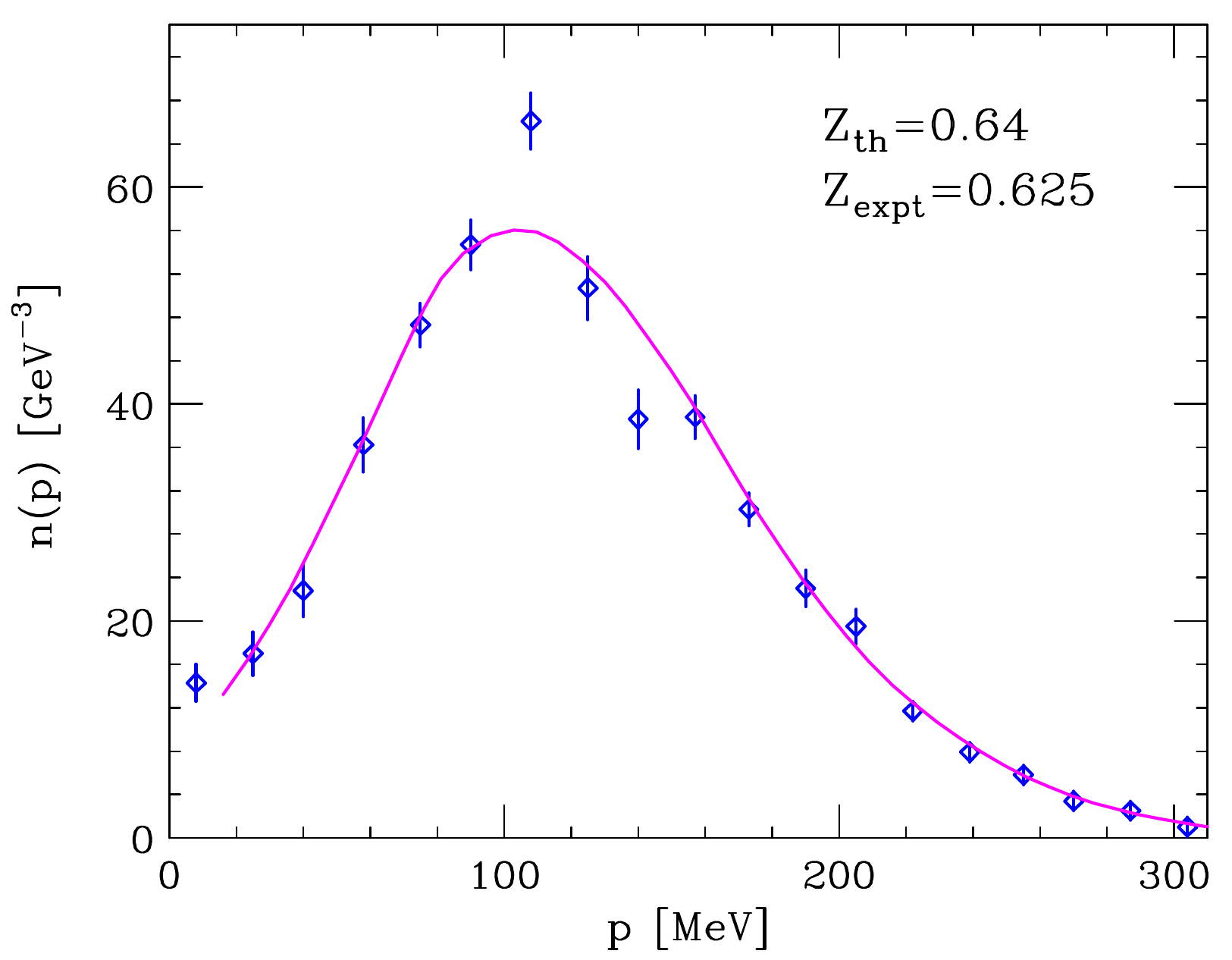}
\end{center}
\caption{Comparison between the momentum distribution of the $p$-state of carbon obtained from the 
measured \isotope[12][]{C}$(e,e^\prime p)$ cross section~\cite{MOUGEY1976461} and the results of a theoretical calculation
performed using the spectral function of Ref.~\cite{LDA}. The corresponding spectroscopic factors are also reported.}
\label{12Ceep}
\end{figure}

Figure~\ref{12Ceep} shows a comparison between the momentum distribution obtained from a measurement of 
the \isotope[12][]{C}$(e,e^\prime,p)$ cross section\textemdash in the kinematic region in which knock out of a $p$-state proton is the dominant reaction mechanism~\cite{MOUGEY1976461}\textemdash and the result of a theoretical calculation 
preformed using the spectral fucntion of Ref.~\cite{LDA}. The corresponding values of the spectroscopic
factors, revealing a $\sim 40$\% missing strength, are also reported. 

The results of Fig.~\ref{12Ceep} clearly imply that the approach based on factorisation and the spectral 
function formalism is best suited to pin down the 1p1h contribution to the cross section of inclusive reactions, 
in which only the outgoing lepton is detected. Note that within the model of Ref.~\cite{nieves}, based on the
mean-field description of the target ground state, this contribution is significantly over-predicted. 

In the superscaling approach of Ref.~\cite{SuSa}, in the other hand,
both the 1p1h contribution and the 2p2h contribution originating from ground-state correlations are 
included in the QE cross section, and cannot be unambiguously identified.

\section{Summary and perspectives}
\label{summary}

The accurate description of neutrino-nucleus interactions in the broad kinematic range relevant to accelerator-based searches of neutrino oscillations will require a combination of the flexibility offered by the factorisation scheme and a treatment of nuclear dynamics beyond the mean-field approximation. Such an approach can be systematically improved to take into account the contribution of 
processes involving MEC, as well as corrections arising from FSI.
The development of a fully consistent framework to perform accurate  calculations of neutrino-nucleus cross sections, will help future neutrino experiments to achieve the required precision in the determination of oscillation parameters. 

The accuracy of a cross-section model can be gauged comparing its prediction to the available electron scattering data.
The approach based on factorisation and realistic spectral functions provides a remarkably good description
of electron scattering processes involving different nuclear targets\textemdash ranging from the few-nucleon systems to nuclei as heavy as iron\textemdash in a 
variety of kinematic conditions. It has to be emphasised that the calculation of electron-nucleus cross sections {\it does not involve any adjustable parameters}. The spectral function is obtained from a nuclear Hamiltonian\textemdash strongly constrained by the observed properties of two- and three-nucleon systems\textemdash and the measured $(e,e^\prime p)$ cross sections, while the description of the primary interaction vertex is based on precisely measured electron-proton and electron-deuteron cross sections. The extension of this treatment to the case of neutrino interactions, 
however, requires additional information on the interaction vertices involving the axial weak current, which have been mainly studied by low-statistic experiments.

The MiniBooNE and K2K Collaborations demonstrated that an increased nucleon axial mass, entailing a modification of the $Q^2$-dependence of the axial form factor, can explain the excess of the measured CCQE cross section with respect to the prediction of the RFGM. On the other hand, the results of theoretical calculations carried out using the canonical value $M_a = 1.03$~MeV led to a widespread consensus that the excess cross section
originates from processes involving MEC. Using the SF formalism in conjunction with the axial form factors recently obtained from lattice calculations~\cite{Park:2021ypf} may greatly help to clarify the role of MEC contributions at fully quantitative level. 

The occurrence of processes with final states involving collective nuclear excitations, defined in Eq.~\eqref{RPA_C}, must be carefully analysed, since the 
corresponding cross section is inherently non factorisable.
The results of the exploratory analysis of Ref.~\cite{Benhar:2009hj} suggest that the kinematic region in which these processes provide important contributions is limited to momentum transfer not exceeding few tens of MeV. In this context, it must be emphasised that, in order to achieve a truly unified treatment of neutrino-nucleus interactions, the regime of low momentum transfer should be described using the same model of nuclear dynamics employed for the description of the IA regime. This point has been recently highlighted by the results reported in Ref.~\cite{Nieves:2017lij}, showing that the RPA contribution predicted by the 
model of Ref.~\cite{nieves} is significantly reduced when dynamical effects not included in the the RFGM are taken into 
account.

Approaches based on factorisation are ideally suited for implementation in available and future neutrino interaction generators. 
An important factor is that new models are correctly implemented in neutrino event generators being those the key to make cross-comparison between different data sets and different nuclei used as the neutrino and anti-neutrino targets. The theoretical framework should allow even generator to be fast. A large ensemble of MC is needed due to the increase statistics available in oscillation experiments, bigger detector and more powerful neutrino beams. Independently from the approach used, the theoretical framework and its computational implementation should be tested and validated on electron scattering data. The high resolution beams and very accurate measurements of electron scattering makes it a key factor in assessing and validating nuclear models that are used in neutrino oscillation and scattering experiments. Electron scattering data will also contribute to determining the accuracy or systematic errors of a particular theoretical model. 
Consistency across theory models building - relativistic vs non-relativistic, different assumptions on final states and final state particles, overlap between reaction mechanisms that dominate at different kinematic regimes - and MC computational implementations are all ingredients and needs that should be satisfied to satisfy the increasing demanding of accuracy from neutrino oscillation experiments.

\backmatter

\bmhead{Acknowledgments}

We acknowledge the outstanding support from the Virginia Tech Physics department, the Virginia Tech Center for Neutrino Physics.
The work of C.M. was supported by the National Science Foundation under award PHYS-2207171 and the U.S. Department of Energy Office of Science under award number DE-SC00023471.

\bibliography{EPJA_review}

\end{document}